# Displaying Asynchronous Reactions to a Document: Two Goals and a Design

*Todd Davies   Benjamin Newman   Brendan O'Connor   Aaron Tam   Leo Perry*
Symbolic Systems Program, Stanford University, Stanford, CA 94305-2150
{davies,newmanb,brendano,aarontam,leoperry}@stanford.edu

## ABSTRACT
We describe and motivate three goals for the screen display of asynchronous text deliberation pertaining to a document: (1) *visibility of relationships* between comments and the text they reference, between different comments, and between group members and the document and discussion, and (2) *distinguishability of boundaries* between contextually related and unrelated text and comments and between individual authors of documents and comments. Interfaces for document-centered discussion generally fail to fulfill one or both of these goals as well as they could. We describe the design of the new version of Deme, a web-based platform for online deliberation, and argue that it achieves the two goals better than other recent designs.

**ACM Classification:** H.5.3 [**Group and Organization Interfaces**]: Asynchronous interaction---Design, Human factors.

## 1. INTRODUCTION
Supporting group interaction around a shared document is challenging for designers of two-dimensional interfaces and asynchronous, text-based groupware. The need to *deliberate (*collaborate, make decisions, or make comments) around documents appears to be one of the main reasons why groups that could otherwise interact virtually and asynchronously via the Internet choose to meet synchronously, either in person or online, often in a richer environment than text-only (e.g. including audio, video, and/or a 3-D environment).

In our usage, a *document* can be in any format, including images, audio and video, but our primary focus here is on digital documents in which most of the information is in text. Discussion that takes place around a document consists of *comments* that pertain either to the document as a whole or to some part of it. The document may be either fixed or evolving as the discussion around it proceeds, but the document assumes an elevated status over the comments made about it, e.g. because it has been chosen for careful discussion, or because its final version will have governing consequences or will represent the outcome of a collaboration. In the latter cases, the group must somehow reach a *decision* relating to the document (e.g. to adopt it).

Distributed asynchronous interaction offers many advantages to groups deliberating about documents: more time for reflection, revision, and information seeking *(cf.* [2])*;* the ability to accommodate people's conflicting schedules; flexible interaction modes through conversion of text to/from speech (e.g. for disabled or less literate users); the easier access, storage, and search afforded by digital archives; and [3] the empowerment of those who are at a disadvantatge when participation involves speaking in a live group. But mapping in-person meetings onto an asynchronous interaction through distributed 2-D text displays entails several types of lost richness, including nonverbal grounding cues [1], spatial depth, the natural use of separate perceptual modalities for document (visual) and discussion (auditory); and the use of a shared temporal progression to guide attention.

## 2. GOALS FOR A DOCUMENT-CENTERED DELIBERATION INTERFACE[1]
The projection of a 3-D, multi-modality, co-located, synchronous deliberation experience onto a 2-D, primarily visual, distributed asynchronous interaction requires essential aspects of face-to-face deliberation to be remapped onto a screen interface. The needed mappings can be judged according to two broad goals:

- **Visible relationshps.** Relationships between comments and the texts they reference, between different comments, and between group members and the document and discussion, should be as visible as possible.

- **Distinguishable boundaries.** Separations between contextually related and unrelated text and comments and between individual authors of documents and comments should be as distinguishable as possible.

Let us consider each of these goals in more detail.

### 2.1 Visible Relationships
Exhibiting relationships between the components of document-centered deliberation (document, comments, and participants) implies a number of refinements of this goal. First, the document text that is the target of deliberation should be *co-visible* (displayed simultaneously) with comments around it, and the identities of comment authors and document text, when relevant, should also be co-visible with their output. Second, the referencing relationship between a comment and its target text should be visible, i.e. the interface should incorporate *ostensive pointing* (meaning that a pointing relationship is displayed on the screen rather than being enacted through a peripheral device) and *in-text placement* of comments. Third, response relationships between comments should be visible through *threading*. And fourth, the reactions of deliberation participants should be visible through *polling and decision features*.

---

[1] This paper focuses on interfaces for visually-abled users. Adapting the analysis presented here for visually impaired users might be possible, but our feeling is that that will require quite a different approach, one we hope to investigate in the future.

## 2.2 Distinguishable Boundaries

Visible relationships can be inadequate, as anyone would know who has used a map with ambiguous place labelings. The interface should also mark boundaries between text that is and is not the reference target of a comment, e.g. though *text highlighting*. Text authored by different people at different times should be distinguishable through *textual boundaries.* The topic of a text should be able to be viewed separately from its main body through *headers*. And obsolete comments (i.e. made on a previous version) should be recognizable through *pertinence markers*.

## 3. ASSESSING DIFFERENT DESIGNS

The principles of visibility and distinguishability can be applied to any interface for document-centered deliberation. Table 1 shows the beginnings of an assessment of text/comment visbility and distinguishability in five popular tools for document-centered discussion.[2] The analysis is far from exhaustive even for this small set of tools – the table is just meant to illustrate how different tools tend to have some but not all of the elements implied by the two principles described above. None of these tools appears to offer integrated decision and polling support-specific features or pertinence markers within the discussion environment, and all of them put users at a serious disadvantage relative to the capabilities of a face-to-face meeting.

**Table 1. Comparison of Document-Deliberation Tools**

| Tool | Visibility | Distinguishability |
|---|---|---|
| D3E | co-visible relations<br>no in-text, pointing | comment boundaries<br>no highlighting |
| Quick Doc Review | in-text comments<br>not co-visible | item demarcation<br>no arbitrary highlights |
| Stet | in-text comments<br>co-visible<br>no pointing | text highlighting<br>no headers |
| Wikimedia | not co-visible<br>no pointing<br>optional threading | no textual boundaries<br>text-comment<br>separation by tabs |
| Microsoft Word | co-visible relations<br>in-text pointing<br>no threading | comment boundaries<br>no headers |

---

[2] D3E is at d3e.sourceforge.net; Quick Doc Review is a product of Quicktopic.com; Stet is under development and visible at gplv3.fsf.org/comments; Wikimedia powers Wikipedia.org; and Word is available through Microsoft.com.

## 4. THE DEME APPROACH

The Deme environment for online deliberation is a tool for document-centered discussion, polling and decision making that incorporates all of the elements derived above from the goals of relational visibility and boundary distinguishability in a new "AJAX" interface, under development.

Figure 1 shows the most recent design of the *meeting area* viewer in Deme. The shaded-in header of a comment in the discussion view pane on the right points to a shaded-in comment reference in the text of a document shown in the item view pane on the left.

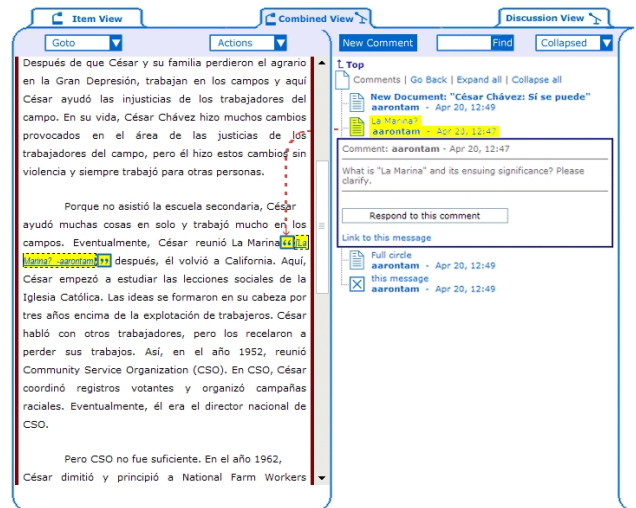

**Figure 1. Document referencing in Deme**

Deme provides co-visibility between document and comments through an optional split-screen view. In-text comment references are transiently pointed to (the dotted-line arrow goes away as soon as the user scrolls) when clicked on, and comments are displayed in the context of hierarchical threads. Members can vote on documents under a variety of decision rules. Boundaries are provided through highlighting, text boundaries, headers, and a versioning system that remembers when comments become obsolete and marks them as such. The design takes advantage of no-page-reload web server calls to provide dynamic relationship visibility and boundary distinguishability.

The principles discussed above led us to incorporate a number of features that are seen in other systems, along with some new ideas such as transient pointing, but in an integrated way that, we argue, puts online deliberators at less of a disadvantage relative to their face-to-face counterparts.